\documentclass[journal=cmatex,manuscript=article]{achemso}

\usepackage[version=3]{mhchem}
\usepackage{graphicx}
\usepackage{multirow}
\usepackage{color}
\usepackage{soul}
\usepackage{amsmath,amssymb}
\usepackage[hidelinks]{hyperref}

\usepackage{makecell}

\usepackage{tablefootnote}

\setkeys{acs}{articletitle = true}
\SectionNumbersOn
\AbstractOn
\title{Ionic  transport in potential coating materials for Mg batteries}

\author{Tina Chen}
\affiliation[UCB]
{Department of Materials Science and Engineering, University of California, Berkeley, Berkeley, CA, United States}
\alsoaffiliation[LBL]
{Materials Science Division, Lawrence Berkeley National Laboratory, Berkeley, CA, United States}

\author{Gopalakrishnan Sai Gautam}
\affiliation[princeton]
{Department of Mechanical and Aerospace Engineering, Princeton University, Princeton, NJ, United States}
\email{gautam91@princeton.edu}

\author{Pieremanuele Canepa}
\affiliation[nus]
{Department of Materials Science and Engineering, The National University of Singapore, 117576, Singapore}
\email{pcanepa@nus.edu.sg}

\begin{document}
%\linenumbers
%%%%%%%%%%%%%%%%%%%%%%%%%%%%%%%%%%%%%%%%%%%%%%%%%%%%%%%%%%%%%%%%%%%%%%%%

%%%%%%%%%%%%%%%%%%%%%%%%%%%%%%%%%%%%%%%%%%%%%%%%%%%%%%%%%%%%%%%%%%%%%%%%
% abstract
%%%%%%%%%%%%%%%%%%%%%%%%%%%%%%%%%%%%%%%%%%%%%%%%%%%%%%%%%%%%%%%%%%%%%%%%
\begin{abstract}
A major bottleneck for the development of Mg batteries is the identification of liquid electrolytes that are simultaneously compatible with the Mg-metal anode and high-voltage cathodes. One strategy to widen the stability windows of current non-aqueous electrolytes is to introduce protective coating materials at the electrodes, where coating materials are required to exhibit swift Mg transport. In this work, we use a combination of first-principles calculations and ion-transport theory to evaluate the migration barriers for nearly  27 Mg-containing binary, ternary, and quaternary compounds spanning a wide chemical space. Combining mobility, electronic band gaps, and stability requirements, we identify MgSiN$_2$, MgI$_2$, MgBr$_2$, MgSe, and MgS as potential coating materials against the highly reductive Mg metal anode, and we find MgAl$_2$O$_4$ and Mg(PO$_3$)$_2$ to be promising materials against high-voltage oxide cathodes (up to $\sim$3 V). 
\end{abstract}
%%%%%%%%%%%%%%%%%%%%%%%%%%%%%%%%%%%%%%%%%%%%%%%%%%%%%%%%%%%%%%%%%%%%%%%%

%%%%%%%%%%%%%%%%%%%%%%%%%%%%%%%%%%%%%%%%%%%%%%%%%%%%%%%%%%%%%%%%%%%%%%%%
\section{Introduction}
\label{sec:intro}
%%%%%%%%%%%%%%%%%%%%%%%%%%%%%%%%%%%%%%%%%%%%%%%%%%%%%%%%%%%%%%%%%%%%%%%%
Mg batteries operating with a Mg-metal anode set a practical alternative to state-of-the-art Li-ion batteries by providing increased volumetric capacities ($\sim$~3833 mAh/cm$^3$ for Mg versus $\sim$~800 mAh/cm$^3$ for Li insertion into graphite) at a lower cost.\cite{Aurbach2000,Yoo2013, Canepa2017} The primary advantage of Mg batteries is the use of the metallic anode since Mg can be safely plated and stripped, dendrite-free, from electrolytic solutions at acceptable current densities ($\sim$0.5--1.0~mA/cm$^2$) \cite{Yoo2013}. While the possibility of plating dendrite-free Mg has been recently challenged by Davidson \emph{et al.}  \cite{Davidson2018} and Ding \emph{et al.} \cite{Ding2018}, the shape of the electrodes employed in the former and the inappropriate classification of globular Mg deposits as dendrites in the latter cast doubt on the formation of Mg dendrites under practical battery conditions.\cite{Ponrouch2019} While Davidson \emph{et al.} observe that sharp edges and/or a rough surface at the Mg electrode can lead to the nucleation and growth of dendrites,\cite{Davidson2018} the potential usage of a metallic, dendrite-free, and safe anode still remains one of the main advantages of Mg-based electrochemical storage.

Nevertheless, a functioning Mg battery is challenged by the poor chemical and electrochemical stability of a handful of liquid electrolytes \cite{Yoo2013, Muldoon2014}, which are compatible with either the Mg-metal anode or a high-voltage cathode, but not with both electrodes simultaneously.  The decomposition of liquid electrolytes on electrodes can lead to the formation of passivating layers that permanently block the electrochemical function of a battery. For example, the formation of a passivating MgO layer on the Mg-metal surface \cite{Aurbach2000, Canepa2017} or at the high-voltage cathode \cite{Ling2015, Ling2017} has lead to rapidly diminishing electrochemical capacities with cycling. Furthermore, such unstable electrolytes can lead to safety issues due to the presence of flammable solvents, as has been observed in Li-ion electrolytes.\cite{Xu2014, MacNeil2002}

One strategy to mitigate the safety challenges of liquid electrolytes is using a less flammable solid material (i.e., a solid electrolyte) capable of transporting Mg ions efficiently. While two examples of such materials exist in the multivalent chemical space, namely, MgSc$_2$X$_4$ (X=S, Se)\cite{Canepa2017a} and ZnPS$_3$,\cite{Martinolich2019} using chalcogenide chemistries to boost multivalent mobility typically correlates with poor electrochemical stabilities and increased electronic conductivity.\cite{Canepa2017a, Wang2019, Canepa2017b}. Note that electronic conductivity within a solid electrolyte is a source of self-discharge and is detrimental to battery performance.

A practical way to solve the inherent liquid/solid electrolyte instability is to develop protective coating materials that can selectively mitigate detrimental decomposition reactions against highly oxidizing cathodes and/or the highly reducing Mg-metal. In practice, the identification of protective coatings consists of identifying a number of materials whose electrochemical stability window (ESW)\cite{Goodenough2010, Xiao2019} is sufficiently large that they span across the HOMO (or LUMO) level of a liquid electrolyte and the chemical potential of the cathode (or anode). Subsequently, candidate coating materials can either be deposited as thin ($\sim$nano-scale) layers via \emph{ex situ} methods or can be formed \emph{in situ} via the reaction of a strategically chosen electrolyte (and additives in liquid electrolytes) and the electrodes.\cite{Chen2010, Myung2005, Aykol2016, Myung2007, Qian2012} 

One potential difference between \emph{ex situ} and \emph{in situ} methods of forming coating materials is the resulting electronic conductivity. In general, a higher electronic conductivity in a coating than the electrolyte is detrimental to the stability of the electrolyte. This is because the drop/gain in chemical potential across the coating may not be sufficient enough to protect the electrolyte from reduction/oxidation.{\cite{Nakamura2019}} Additionally, a thicker coating layer accommodates a higher chemical potential difference and becomes more suitable for accommodating an electrolyte with a small ESW. Hence, the choice and thickness of a coating (and its electronic conductivity) can be calibrated depending on the intrinsic electronic conductivity of the electrolyte in \emph{ex situ} methods. However, if a coating is obtained via \emph{in situ} reactions at the electrode$||$electrolyte interface, a careful analysis of the properties (ESW and electronic conductivity) of the phase formed at the interface will be necessary to ensure that the battery doesn't exhibit self-discharge. Nevertheless, a candidate coating is optimal if it exhibits a sufficiently large ESW and a significantly low electronic conductivity.

Recently,\cite{Chen2019} we reported the ESWs of several Mg-containing compounds, which can be potential coating materials for Mg batteries. We estimated the ESWs via the construction of grand-potential phase diagrams based on density functional theory (DFT) calculations.\cite{Richards2015} From the calculated ESWs, we identified binary Mg-halides (MgF$_2$, MgCl$_2$, MgBr$_2$ and MgI$_2$), and Mg(BH$_4$)$_2$ as possible anode coating candidates and MgF$_2$, Mg(PO$_3$)$_2$, and MgP$_4$O$_{11}$ as potential cathode coatings. Analogous work by Snydacker \emph{et al.}\cite{Snydacker2017} also proposed a subset of the aforementioned materials. However, the utilization of these proposed materials as effective coatings demands that Mg$^{2+}$ transport in their structures is facile under battery operating conditions. Hence, it is paramount to evaluate the Mg mobility on any candidate coating. 

Using first-principles calculations, we systematically assess the barriers and band gaps for Mg migration in a total of 27 candidate coating materials. In addition to the compositions listed above, we also considered materials that are stable against Mg metal\cite{Chen2019,Snydacker2017} (potential anode coatings) and analogous chemistries that have been employed in Li-systems (e.g., Li-Nb oxides).\cite{Chen2010, Myung2005, Aykol2016, Myung2007, Qian2012} Although prior studies have demonstrated \cite{Tchitchekova2018, Emly2015, Kolli2018, Ling2017a} that the lack of Mg (or multivalent) mobility in several structures relates to a combination of stronger electrostatic interactions of a 2+ charge with its surrounding anion environment (versus 1+ charge of monovalent ions) and strong coordination preferences \cite{Rong2015,Xiao2019, Richards2015, Lacivita2019, Tang2017, Zhu2015}, Mg mobility has not been rigorously quantified yet for potential coating chemistries. 

Since the range of thickness in coating materials is often limited to a few tens of nanometres (0.5--50 nm),\cite{Culver2019} the stringent criteria for ionic mobility required in cathodes or (solid) electrolytes can be slightly relaxed in coating materials. Thus, we used a range of maximum Mg migration barriers, namely $\sim$600-980~meV, to identify candidates that can operate under a variety of battery conditions (see Section~\ref{sec:thresholds}). Based on these limits and the calculated migration barriers, we have identified MgSiN$_2$, MgI$_2$, MgBr$_2$, MgSe, and MgS  as potential anode coatings and MgAl$_2$O$_4$ and Mg(PO$_3$)$_2$ as potential cathode coatings. We also analyzed the migration topology in a set of candidate coatings and found significant similarities to topologies observed in ternary Mg oxides. Finally, our work will offer useful guidance in understanding the electrochemical stability, Mg mobility, and electronic properties in several Mg-containing compounds and enable the development of practical Mg batteries.

%%%%%%%%%%%%%%%%%%%%%%%%%%%%%%%%%%%%%%%%%%%%%%%%%%%%%%%%%%%%%%%%%%%%%%%%

%%%%%%%%%%%%%%%%%%%%%%%%%%%%%%%%%%%%%%%%%%%%%%%%%%%%%%%%%%%%%%%%%%%%%%%%
\section{Migration barrier thresholds in coating materials for Mg batteries}
\label{sec:thresholds}
%%%%%%%%%%%%%%%%%%%%%%%%%%%%%%%%%%%%%%%%%%%%%%%%%%%%%%%%%%%%%%%%%%%%%%%%

The assessment of microscopic migration barriers to evaluate the ability of macroscopic Mg transport in coating materials becomes relevant only if the calculated barriers can be connected to macroscopic properties, such as diffusion coefficients, $D$. Using the Arrhenius expression of Eq.~\ref{eq:diffusivity}, we can estimate the diffusivity of a given ion (e.g., Mg$^{2+}$) in a solid, given a barrier along a microscopic (or local) migration pathway ($E_m$).
\begin{equation}
 D = f a^2 \nu \exp{\left(-\frac{E_m}{k_B T}\right)}
\label{eq:diffusivity}
\end{equation}
where $a$, $\nu$, $f$, $k_B$, and $T$ are the hopping distance along a migration pathway, vibrational frequency of Mg in a host structure, correlation factor, Boltzmann constant, and temperature, respectively. Typically, $a$ and $\nu$ are of the order of $\sim$3~\AA{} and $10^{12}$~Hz, respectively, and do not vary significantly in most solids.\cite{Ven2001} Thus, the  governing variable for $D$ in Eq.~\ref{eq:diffusivity} is $E_m$, which is a chemistry (oxides vs.\ sulfides), structure (layered vs.\ spinel), and pathway (tetrahedral $\rightarrow$ octahedral $\rightarrow$ tetrahedral or octahedral $\rightarrow$ tetrahedral $\rightarrow$ octahedral) dependent property. Additionally, we only consider local migration pathways that form percolating networks,\cite{Ven2001,Gautam2017} i.e., pathways that are sufficiently connected through the lattice, enabling Mg to diffuse from one end of the lattice to the other along at least one crystallographic direction.

We assume that ionic diffusion follows a random-walk behavior without any long-range correlation effects between Mg sites, i.e., $f \sim 1$ in Eq.~\ref{eq:diffusivity} . Further, the diffusion length $l$ of Mg$^{2+}$ across a coating layer (of thickness $\sim l$) scales as,
\begin{equation}
l = \sqrt{D t}
\label{eq:lenght}
\end{equation}
with $t$ the time for (dis)charge, i.e., the time taken for Mg$^{2+}$ to diffuse through the coating layer. By fixing the time $t$ to (dis)charge a battery, at a given thickness of coating layer (i.e., the diffusion length), one arrives at a minimum required Mg-diffusivity, $D_{\mathrm{min}}$, via Eq.~\ref{eq:lenght}, which is equivalent to a maximum tolerable migration barrier (E$_{m}^{\mathrm{max}}$) from Eq.~\ref{eq:diffusivity}. Thus, potential candidates are those that exhibit values of E$_{m}$ lower than the E$_{m}^{\mathrm{max}}$.

Figure 1 illustrates mobility considerations, which sets general guidelines, apart from thermodynamic stability, for material selection in batteries (including Li- and Na-systems), which we apply to identify potential Mg coatings. Figure 1 plots E$_{m}^{\mathrm{max}}$ at various thicknesses (or equivalently particle sizes), which sets $l$, and (dis)charge rates of potential coating materials (or cathode) \cite{Canepa2017, Rong2015} that determine $t$. While changes in temperature affect the value of $D_{\mathrm{min}}$ (and E$_{m}^{\mathrm{max}}$), typical Mg batteries are cycled at 60~$^{\circ}$C to mitigate the poor kinetics of Mg$^{2+}$ diffusion \cite{Aurbach2000, Sun2016}. Hence, we include both room temperature (orange bar, Figure 1) and 60~$^{\circ}$C (green bar) to estimate E$_{m}^{\mathrm{max}}$. 

\begin{figure}[ht]
\includegraphics[width=\columnwidth]{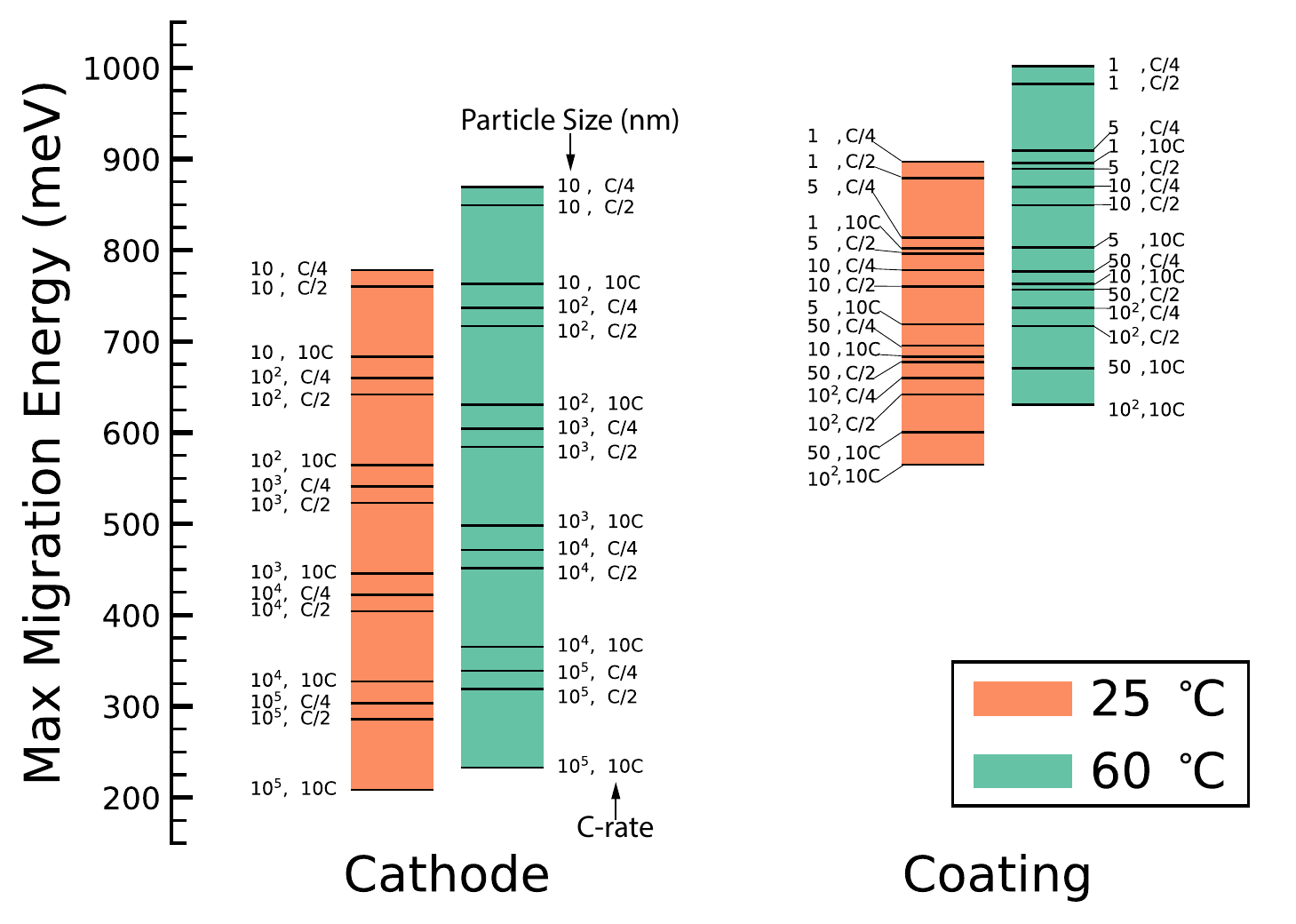} 
\caption{
Maximum tolerable migration barriers (E$_{m}^{\mathrm{max}}$) in cathode and coating materials for different sizes of cathode particles and different thicknesses of coating layers, operating at different (dis)charge rates (in units of C-rate) and at two different temperatures (25 and 60~$^{\circ}$C). The particle sizes/layer thicknesses are given in units of nm.
\label{fig:thickness}
}
\end{figure}

In Figure~\ref{fig:thickness},  we consider rates of (dis)charge for battery operations ranging from extremely fast (dis)charging at 10C (i.e., 6 minutes) to C/4 (4 hours). Typically, slow (dis)charge regimes are closer to the conditions  of thermodynamic equilibrium and are used for laboratory-scale experiments, while fast charging{\cite{Ahmed2017_fast}} has become more prevalent in portable electronics recently. Consequently, the combination of low (dis)charge rates and high temperatures allow the operation of coating materials (and cathodes) with higher migration barriers. For example, a coating layer of thickness $\sim$50~nm operating at 25~$^{\circ}$C and 10C can tolerate  E$_{m}^{\mathrm{max}}$ of $\sim$600~meV, while a similarly thick coating layer at 60~$^{\circ}$C and C/4 can accommodate E$_{m}^{\mathrm{max}}$ $\sim$777~meV.

Cathode particles in Li-ion batteries are routinely coated with layers ranging in thickness between 1 and 50~nm.\cite{Ohta2007, Myung2005} Thus, the thickness of coating layers is normally 1-2 orders of magnitude lower than that of cathode particles,\cite{Myung2005, Aykol2016, Myung2007, Chen2010,Qian2012,Culver2019} which is reflected in a higher tolerable E$_{m}^{\mathrm{max}}$ for coatings compared to cathodes (Figure \ref{fig:thickness}). Coating layer thickness can also vary depending on the synthesis and processing techniques utilized. For example, atomic layer deposition (ALD) can produce coatings thinner than a nanometer while pulsed laser deposition (PLD) typically yields thicker coatings (50-1000 nm).\cite{Culver2019} Hence, given a variety of operating conditions that require different E$_m^{\mathrm{max}}$ for coatings (Figure~\ref{fig:thickness}), we select two specific cases to represent ``reasonable'', yet significantly different, battery operating conditions: i) 50~nm coating thickness at 25~$^{\circ}$C and 10C which yields E$_{m}^{\mathrm{max}}$ of $\sim$~600~meV, and ii) 1~nm coating thickness at 60~$^{\circ}$C and C/2 resulting in E$_{m}^{\mathrm{max}}$ of $\sim$~980~meV. Nevertheless, a thermodynamically stable coating material with a low migration barrier ($<$ 400 meV) will always be ideal. 

%%%%%%%%%%%%%%%%%%%%%%%%%%%%%%%%%%%%%%%%%%%%%%%%%%%%%%%%%%%%%%%%%%%%%%%%
\section{Methods}
\label{sec:methods}
%%%%%%%%%%%%%%%%%%%%%%%%%%%%%%%%%%%%%%%%%%%%%%%%%%%%%%%%%%%%%%%%%%%%%%%%

\subsection{First-principles calculations}
\label{subsec:firs_princ}

To calculate the E$_{m}$ for a migration pathway in a given structure of a candidate coating material, we employ the nudged elastic band (NEB) method\cite{Henkelman2000, Sheppard2008} coupled with density functional theory (DFT) calculations,\cite{Hohenberg1964, Kohn1965} as implemented in the Vienna ab initio simulation package (VASP).\cite{Kresse1993, Kresse1996} The exchange and correlation energy are approximated with the Perdew-Burke-Ernzerhof (PBE)\cite{Perdew1996} functional. The total energy is sampled on a well-converged $k$-point mesh with a grid density of 1000/atom together with the projector-augmented wave (PAW) theory\cite{Kresse1999} and a 520 eV plane-wave cut-off for the valence electrons. Unless otherwise mentioned, for each compound considered, we evaluate E$_m$ in its ground state configuration as per the Materials Project (MP) database.\cite{Jain2013} Supercells used for NEB calculations introduce a minimum distance of at least 8~\AA{} between the migrating Mg ions to minimize fictitious interactions across periodic boundaries. The total energy is converged within 10$^{-5}$~eV per supercell. The end-point structures, i.e., the initial and final states along the Mg migration pathway, are fully relaxed until the forces on the atoms converged within 10$^{-2}$~eV/\AA{}, whereas the NEB forces are converged within 0.05~eV/\AA{}. Unless noted differently, seven distinct images are used between the endpoints to evaluate the ion migration trajectory. 

Band-gaps of the materials studied are evaluated from the density of states (DOS)  computed with the hybrid (25\% Hartree-Fock) range-separated exchange-correlation functional, HSE06.\cite{Heyd2003, Heyd2006, Krukau2006} For DOS calculations, we fully relaxed coordinates, cell, and shape of the bulk unit cells, in accordance with the input settings used in the MP.\cite{Jain2011} Notably, the band gap predictions with HSE06 shows good agreement with experiments and/or other theoretical calculations. For example, our HSE06 calculated band gaps for Mg$_2$Si ($\sim$~0.83~eV) and Mg$_2$Ge ($\sim$~0.61~eV) are similar to previous studies, namely $\sim$~0.77-0.8~eV for Mg$_2$Si and $\sim$~0.67-0.74~eV for Mg$_2$Ge.\cite{AU-YANG1969}

\subsection{Challenges in computing migration barriers in coating materials}

The DFT+NEB framework has been employed to evaluate E$_ {m}$ for a wide variety of electrode structures\cite{Meutzner2018, Canepa2017} containing open-shell transition metals, owing to lower computational costs compared to \emph{ab initio} molecular dynamics (AIMD) simulations and hence represents the standard to evaluate ionic migration barriers in solids. However, a number of challenges remain when the DFT+NEB methodology is applied to materials containing closed-shell non-redox-active (transition) metals, as in the case of coating materials. 

Typically, migration barriers are evaluated in two limits of Mg$^{2+}$ concentration: \emph{i}) low vacancy limit, where the barrier for the migration of an isolated Mg-vacancy is evaluated, and \emph{ii}) high vacancy limit, i.e., migration barrier of an isolated Mg$^{2+}$. In both scenarios, Mg-ions hop via a site-vacancy (or an interstitial-based) mechanism, where the diffusion carrier (Mg-vacancy or interstitial) is introduced and the migration barrier of the carrier is evaluated on all possible paths that can give rise to long-range Mg diffusion. We assess Mg migration in the low-vacancy limit, i.e., one Mg vacancy per supercell, for all the coating materials considered in this work. 

To maintain the charge-neutrality of the underlying structure upon introducing the Mg-vacancy (or Mg atom), the valence electrons of the removed (added) Mg atom must be reintroduced (removed) in the simulation cell, i.e., the charge imbalance created by a local Mg diffusion carrier should be compensated by the addition/removal of electrons. For semiconducting/insulating cathode materials that contain open-shell redox-active species (e.g., 3\emph{d}/4\emph{d} transition metals), the addition (removal) of electrons is accommodated by a change in the oxidation state of a ``nearby'' redox-active species. In the case of metallic electrodes, local charge imbalances are efficiently screened by the metal electron density and do not have to be explicitly accounted for in DFT calculations. 

However, the addition/removal of electrons with the creation of local Mg-carriers is required for candidate coatings in DFT calculations since most of the candidates: $i$) are highly stoichiometric and have elements with well-defined oxidation states, $ii$) do not contain any redox-active species, and $iii$) are non-metallic. Thus, modelling Mg migration in potential coatings requires ``charge compensation". Since NEB calculations are performed with periodic boundary conditions, the additional electrons for charge-compensation are introduced as a homogeneous background charge, termed ``jellium". To minimize the number of additional electrons that contribute to the jellium background compensating one Mg vacancy, we used the soft-core Mg PAW potential with only two valence electrons [$s^2p^0$]  in our DFT+NEB calculations. Note that in a practical battery device, local charge imbalances in non-metallic coatings are usually compensated either by Mg-atom transfer at the coating$||$electrode interface or through intrinsic defects already present in the material.

\subsection{Workflow to assess Mg migration barriers}

\begin{figure}[!ht]
\includegraphics[width=\columnwidth]{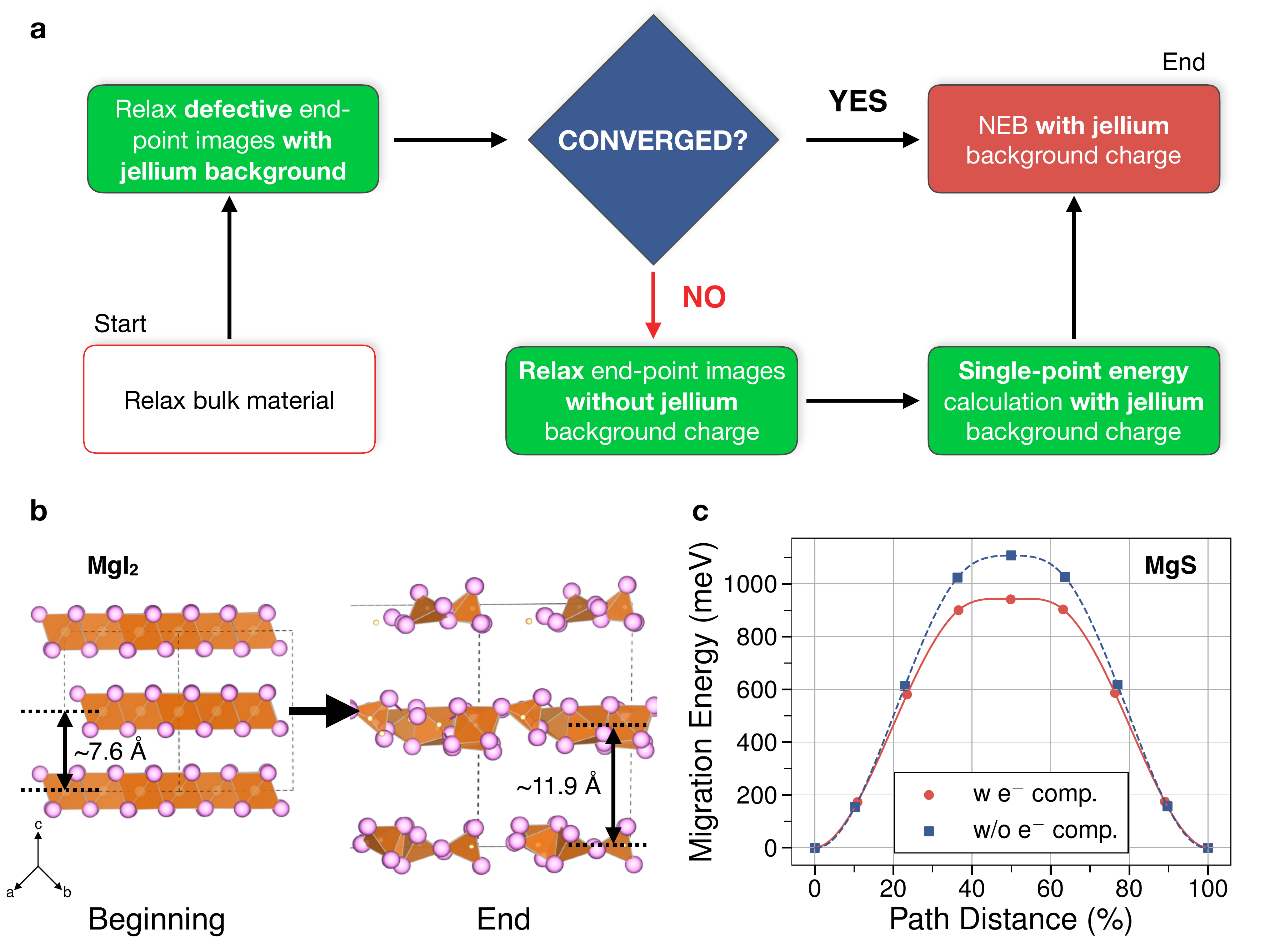}
\caption{{\bf a} Flow-chart to accurately compute migration barriers in charge-compensated candidate coating (and solid electrolyte) systems. {\bf b} Example of failed convergence of end-point images in MgI$_2$ in the presence of a jellium background resulting in significant distortion of layers of edge-sharing MgI$_6$. The structural distortion is highlighted by the change in interlayer distance before ($\sim$7.6~\AA) and after relaxation ($\sim$11.9~\AA). {\bf c} Comparison of migration energies for Mg ions in MgS after the structure relaxation of the end-point images with jellium background charge (red dots) and without jellium (blue squares).
\label{fig:flow_chart}
}
\end{figure}

While previous reports have used the jellium background for charge compensation and computed migration barriers, the accuracy of this approach has not been reliably tested.\cite{Wang2015, Xiao2019, Tang2017, He2017} In particular, it is challenging for any self-consistent procedure, such as DFT with a semi-local PBE functional, to appropriately localize the jellium charge, leading to problematic convergence of the charge density, the total energy, and the atomic forces. Poor convergence of the charge density can eventually cause appreciable deformation of the underlying structure, such as significant rearrangement of atomic positions and/or large changes in volume or shape, resulting in an inability to accurately assess migration barriers via the NEB method. In our work, we encountered significant structural distortions owing to poorly converged charge density in the MgX$_2$ (X = Cl, Br, and I) structures, where layers of edge-sharing MgX$_6$ octahedra are highly distorted upon structure relaxation in the presence of a jellium background, as depicted for MgI$_2$ in Figure~{\ref{fig:flow_chart}}{\bf b}.

To ensure that our structural model retains charge neutrality while not exhibiting major structural distortions, we adopt the workflow of Figure~\ref{fig:flow_chart}{\bf a} to compute the migration barriers. This strategy can be implemented whenever the introduction of a compensating background charge leads to unphysical structures.

The workflow of Figure~\ref{fig:flow_chart}{\bf a} is as follows: 
\begin{enumerate}
    \item Relax (coordinate, shape, and volume) the charge-neutral bulk material with DFT computational settings (see Section~\ref{subsec:firs_princ}).
    \item Relax (coordinate, shape, and volume) the end-point structures, including one or multiple diffusing carriers (Mg-vacancies in our case). In this step, the vacancies of the migration species are explicitly charge-compensated with the jellium background.
    \item Verify convergence of end-point geometries. Specifically, ensure that the relaxed end-point geometries are not significantly different from the starting structure. If the end-points are satisfactorily converged, proceed to step 4, else proceed to step 5.
    \item Perform a NEB where each interpolated site (or image) along the elastic band is charge-compensated, similar to the end-points, and extract migration barrier.
    \item Relax (coordinate, shape, and volume) the end-points without charge-compensation, ensuring that the relaxed geometries do not exhibit significant distortions and proceed to step 6. If the relaxed end-points are significantly perturbed from the starting geometries, accurate migration barriers can only be assessed using more computationally expensive techniques (e.g., AIMD) and/or by using a better functional to describe the electronic exchange-correlation (e.g., hybrid functionals). 
    \item Using relaxed geometries from step 5, perform a single self-consistent field calculation for the end-points incorporating charge-compensation.
    \item Using the relaxed geometries of step 5, construct the elastic band and perform a NEB calculation including charge-compensation. The barrier is extracted using the total energies of the end-points from step 6.
\end{enumerate}

In systems that do not exhibit significant structural distortion when charge-compensation is included in the end-point relaxation calculation, e.g., MgS after step 3, the migration barrier evaluated using step 4 ($\sim$941~meV, red line in Figure~{\ref{fig:flow_chart}}{\bf c}) is significantly lower ($\sim$160~meV) compared to that obtained using steps 5-7 ($\sim$1107~meV, blue line in Figure~{\ref{fig:flow_chart}}{\bf c}). Hence, we expect the barriers evaluated using steps 5-7 for MgX$_2$ (X~=~Cl, Br, and I) structures to be over(under)estimated by $\sim \pm$160~meV, which is equivalent to approximately three orders of magnitude difference in diffusivity (from Eq.~\ref{eq:diffusivity}). Unless explicitly mentioned, all systems other than MgX$_2$ (X~=~Cl, Br, and I) are investigated using steps 1-4, with an accuracy of $\sim \pm$50~meV for the DFT+NEB framework.\cite{Liu2015,Rong2015}

%%%%%%%%%%%%%%%%%%%%%%%%%%%%%%%%%%%%%%%%%%%%%%%%%%%%%%%%%%%%%%%%%%%%%%%%
\section{Results}
\label{sec:results}
%%%%%%%%%%%%%%%%%%%%%%%%%%%%%%%%%%%%%%%%%%%%%%%%%%%%%%%%%%%%%%%%%%%%%%%%

\subsection{Mg migration barriers in candidate materials}

\begin{figure*}[!ht]
\includegraphics[width=1\textwidth]{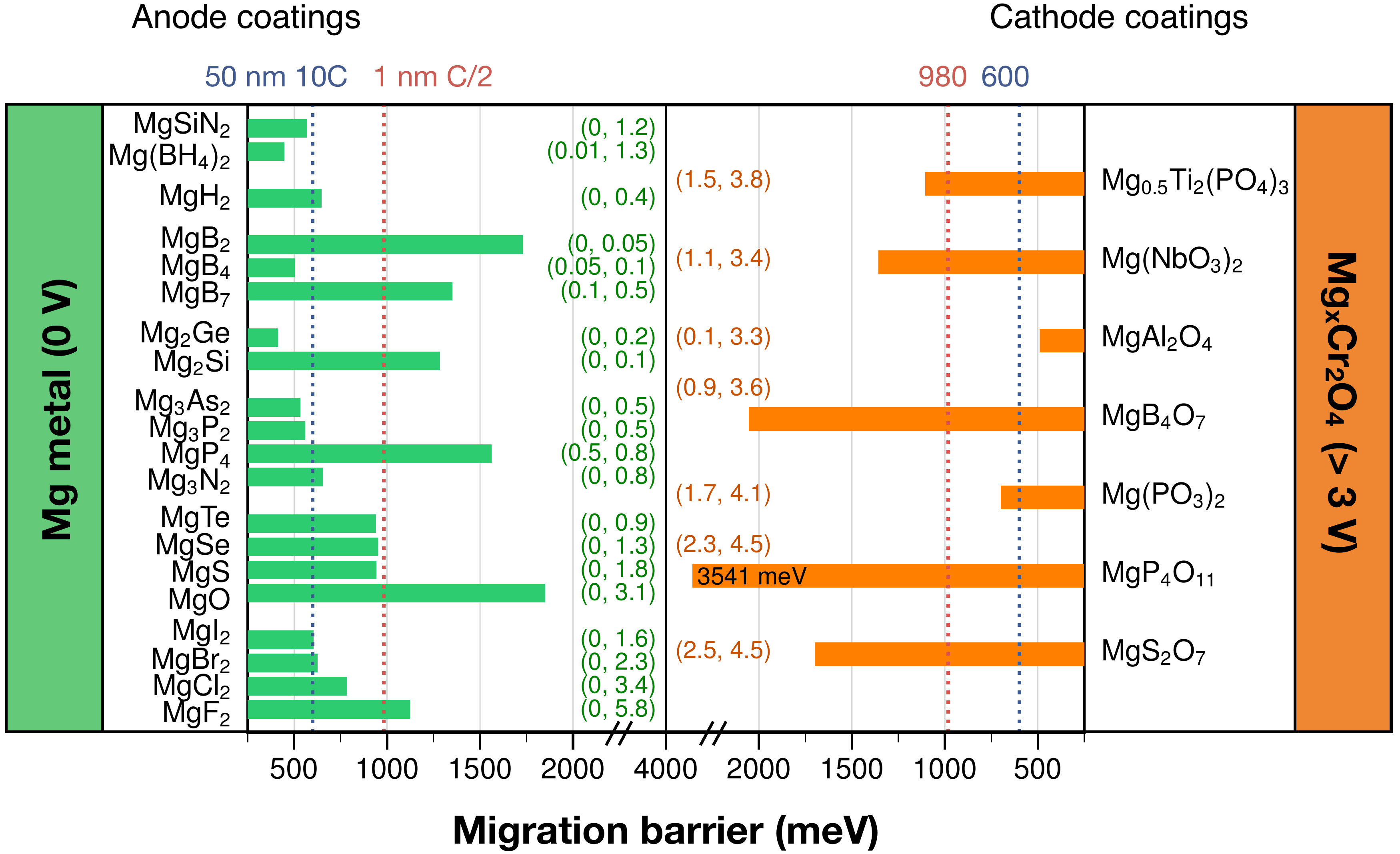}
\caption{Calculated migration barriers for anode (left) and cathode (right) coating candidates. Dotted vertical lines indicate strict ($\sim$600 meV, blue) and lenient ($\sim$980 meV, red) E$_{m}^{\mathrm{max}}$ values (see Section~{\ref{sec:thresholds}}). The numbers in brackets, provide the (reductive, oxidative) stability limits, in V vs. Mg metal, as reported by Chen \emph{et al.}\cite{Chen2019} The E$_m$ for Mg(BH$_4$)$_2$ should exhibit a higher error than other materials due to challenges in converging the NEB.
\label{fig:barriers_bar}
}
\end{figure*}

Figure \ref{fig:barriers_bar} shows the calculated migration barriers for potential binary, ternary and quaternary coating materials for anodes (left, green bars) and cathodes (right, orange bars), as well as their calculated reductive and oxidative stabilities (numbers in parenthesis, in units of V vs.~Mg metal). The computed migration barriers are compared against E$_{m}^{\mathrm{max}}$ values (see Section~\ref{sec:thresholds}) of $\sim$600~meV (50~nm+25~$^{\circ}$C+10C) and $\sim$~980~meV (1~nm+60~$^{\circ}$C+C/2), signifying ``strict'' and ``lenient'' mobility specifications, respectively. The calculated E$_m$, band gaps (this work) and ESWs (from Chen \emph{et al.}\cite{Chen2019}) are also reported in Table~\ref{tbl:summary}. Variation of the calculated migration energies along the migration paths of all materials listed in Table~{\ref{tbl:summary}} are provided in Section~S1 of the supporting information (SI). We also evaluated E$_m$ in a set of metastable polymorphs at the compositions of Mg$_3$P$_2$, Mg$_2$Si, MgSe$_2$, and MgTe$_2$, which are displayed in Section~S2 of the SI.

There are several materials that show appropriate Mg mobility under the strict 600~meV criterion, namely MgSiN$_2$ (570~meV), MgB$_4$ (504~meV), Mg$_2$Ge (414~meV), Mg$_3$As$_2$ (534~meV), and Mg$_3$P$_2$ (560~meV) on the anode and MgAl$_2$O$_4$ (491~meV) on the cathode. Unsurprisingly, using a lenient threshold of 980~meV, we obtain additional candidates, including MgH$_2$ (647~meV), Mg$_3$N$_2$ (655~meV), MgTe (939~meV), MgSe (950~meV), MgS (943~meV), MgI$_2$ (604$\pm$160~meV), MgBr$_2$ (627$\pm$160~meV), and MgCl$_2$ (786$\pm$160~meV) on the anode and Mg(PO$_3$)$_2$ (699~meV) on the cathode. Although we report a low E$_m$ for Mg(BH$_4$)$_2$ ($\sim$448 meV), we encountered significant challenges in converging the NEB for this material. For example, the ground state structure of Mg(BH$_4$)$_2$ in Materials Project (ID: mp-1200811 and space group $Ia\overline{3}d$) contains 264 atoms within its unit cell, making the computational cost of the NEB calculation prohibitive. Despite using a Mg(BH$_4$)$_2$ polymorph with a smaller unit cell (mp-1192265, 22 atoms and $P\bar{4}n2$ space group), we could converge the elastic band only by using a significantly higher force threshold ($\sim$~0.13~eV/\AA{}), which increases the error associated with the reported E$_m$. 

Considering the strict E$_{m}^{\mathrm{max}}$, only MgSiN$_2$ has an ESW $>$ 1~V among the anode coatings of Figure~3, with an oxidative stability up to 1.2~V vs.\ Mg metal. Under lenient operating conditions, MgI$_2$, MgBr$_2$, and MgCl$_2$ also become promising anode coatings, with oxidative stabilities up to 1.2~V, 1.6~V, and 3.4~V, respectively. On the cathode side, MgAl$_2$O$_4$ has a large ESW of 3.2 V, with a reductive stability of $\sim$0.1~V vs.\ Mg. However, the oxidative stability of MgAl$_2$O$_4$ is only up to 3.3 V, which indicates that MgAl$_2$O$_4$ may not be compatible vs. higher voltage oxide cathodes such as Mg$_{\rm x}$Cr$_2$O$_4$ ($\sim$3.5 V)\cite{Chen2017} but can still be compatible with lower voltage oxide cathodes, such as Mg$_{\rm x}$V$_2$O$_5$ ($\sim$2.5~V)\cite{Gautam2015,Jordi2019_V2O5} and Mg$_{\rm x}$Mn$_2$O$_4$ ($\sim$2.8~V).\cite{Gautam2017,Hannah2018,Kim2015a} On the other hand, Mg(PO$_3$)$_2$, which becomes viable under a lenient E$_{m}^{\mathrm{max}}$, exhibits an ESW of 2.4~V with an oxidative stability up to 4.1~V, signifying compatibility with higher-voltage Mg cathodes.  In addition, MgSiN$_2$, MgAl$_2$O$_4$, and Mg(PO$_3$)$_2$ exhibit large band gaps ($>$5~eV) and are good electronic insulators.

In general, there is no correlation between electronic band gaps and Mg migration barriers (Table~\ref{tbl:summary}) across chemistries and structures. For example, both MgB$_2$ and MgB$_4$O$_7$ display significant migration barriers ($>$1700~meV) while possessing contrasting band gaps of 0.1 and 6.91 eV, respectively. However, within an anion group (e.g., chalcogens), there is a direct correlation between lower band gaps and lower migration barriers. For example, the migration barrier in Mg-chalcogenides varies as MgO (1851~meV) $>$ MgSe (950~meV) $>$ MgS (943~meV) $>$ MgTe (939~meV), identical to the variation in band gaps, MgO (6.02 eV) $>$ MgSe (3.00 eV) $>$ MgS (2.94 eV) $>$ MgTe (2.79 eV). An analogous correlation can also be found in Mg-pnictides and layered Mg-halides, with barriers and band gaps varying as Mg$_3$N$_2$ $>$ Mg$_3$P$_2$ $>$ Mg$_3$As$_2$ and MgCl$_2$ $>$ MgBr$_2$ $>$ MgI$_2$ respectively. The decrease in both band gaps and migration barriers moving down the anion group (Cl $\rightarrow$ Br $\rightarrow$ I) can be attributed to the larger volume per anion within the structure and the weaker overlap between atomic orbitals. 

\begin{table*}
  \caption{Computed migration energy (in meV) of Mg$^{2+}$ ions in structures considered. The Materials Project mp-IDs or the collection codes from the inorganic crystal structure database (ICSD) are listed for each structure. The ESWs (V vs.\ Mg), as reported in Ref. \citenum{Chen2019} and the computed electronic band gaps (in eV) are also listed. Unless explicitly mentioned, the overall accuracy of our migration barriers is $\pm$ 50~meV.}
  \label{tbl:summary}
  \begin{tabular*}{\textwidth}{@{\extracolsep{\fill}}llcccc@{}}
    \hline \hline
    {\bf Material} &{\bf MP/ICSD} & {\bf ESW} & {\bf Band Gap} & {\bf E$_\mathbf{m}$} \\
    \hline
    Mg$_3$N$_2$ & mp-1559 & 0.80 & 3.39 & 655 \\ 
    MgSiN$_2$ & mp-3677 & 1.20 & 5.35 & 570 \\
    Mg(BH$_4$)$_2$ & mp-1192265 & 1.25 & 6.35 & 448\textsuperscript{\emph{a}} 
 \\
    MgH$_2$ & mp-23711 & 0.42 & 4.05 & 647\\
    MgB$_2$ & mp-763 & 0.05 & 0.10 & 1729\\
    MgB$_4$ & mp-365 & 0.07 & 0.03 & 504\\
    MgB$_7$ & mp-978275 & 0.41 & 1.73 & 1352\\
    Mg$_2$Ge & mp-408 & 0.21 & 0.61 & 414\\
    Mg$_2$Si & mp-1367 & 0.12 & 0.83 & 1284\\
    Mg$_3$As$_2$ & mp-1990 & 0.52 & 2.23 & 534\\
    Mg$_3$P$_2$ & mp-2514 & 0.52 & 2.43 & 560\\
    MgP$_4$ & mp-384 & 0.27 & 0.69 & 1563\\
    MgO & mp-1265 & 3.08 & 6.02 & 1851\\
    MgS & mp-1315 & 3.39 & 2.94 & 943\\
    MgSe & mp-1018040 & 1.26 & 3.00 & 950\\
    MgTe & mp-1039 & 0.88 & 2.79 & 939\\
    MgF$_2$ & mp-1249 & 5.78 & 8.70 & 1123\\
    MgCl$_2$ & mp-23210 & 3.39 & 6.84 & 786$\pm$160\\
    MgBr$_2$ & mp-30034 & 2.28 & 4.90 & 627$\pm$160\\
    MgI$_2$ & mp-23205 & 1.60 & 3.13 & 604$\pm$160\\
    Mg$_{0.5}$Ti$_2$(PO$_4$)$_3$ & mp-1043685 & 3.55 & 4.07 & 1105\\
    MgNb$_2$O$_6$ & mp-17953 & 2.29 & 4.46 & 1358\\
    MgAl$_2$O$_4$ & mp-3536 & 3.13 & 6.77 & 491\\
    MgB$_4$O$_7$ & mp-14234 & 2.74 & 6.91 & 2056\\
    Mg(PO$_3$)$_2$ & mp-18620 & 2.33 & 7.11 & 699\\
    MgP$_4$O$_{11}$ & mp-15437 & 2.24 & 6.70 & 3541\\
    MgS$_2$O$_7$ & ICSD 426707 & 1.93 & 7.19 & 1699\\
    \hline \hline
  \end{tabular*}
  \begin{flushleft}
  \textsuperscript{\emph{a}}{\small Reported barrier should exhibit error higher than $\pm$~50~meV.} 
  \end{flushleft}
  \end{table*}

\subsection{Mg migration topology of selected coating materials}

\begin{figure}[!ht]
\includegraphics[width=0.8\columnwidth]{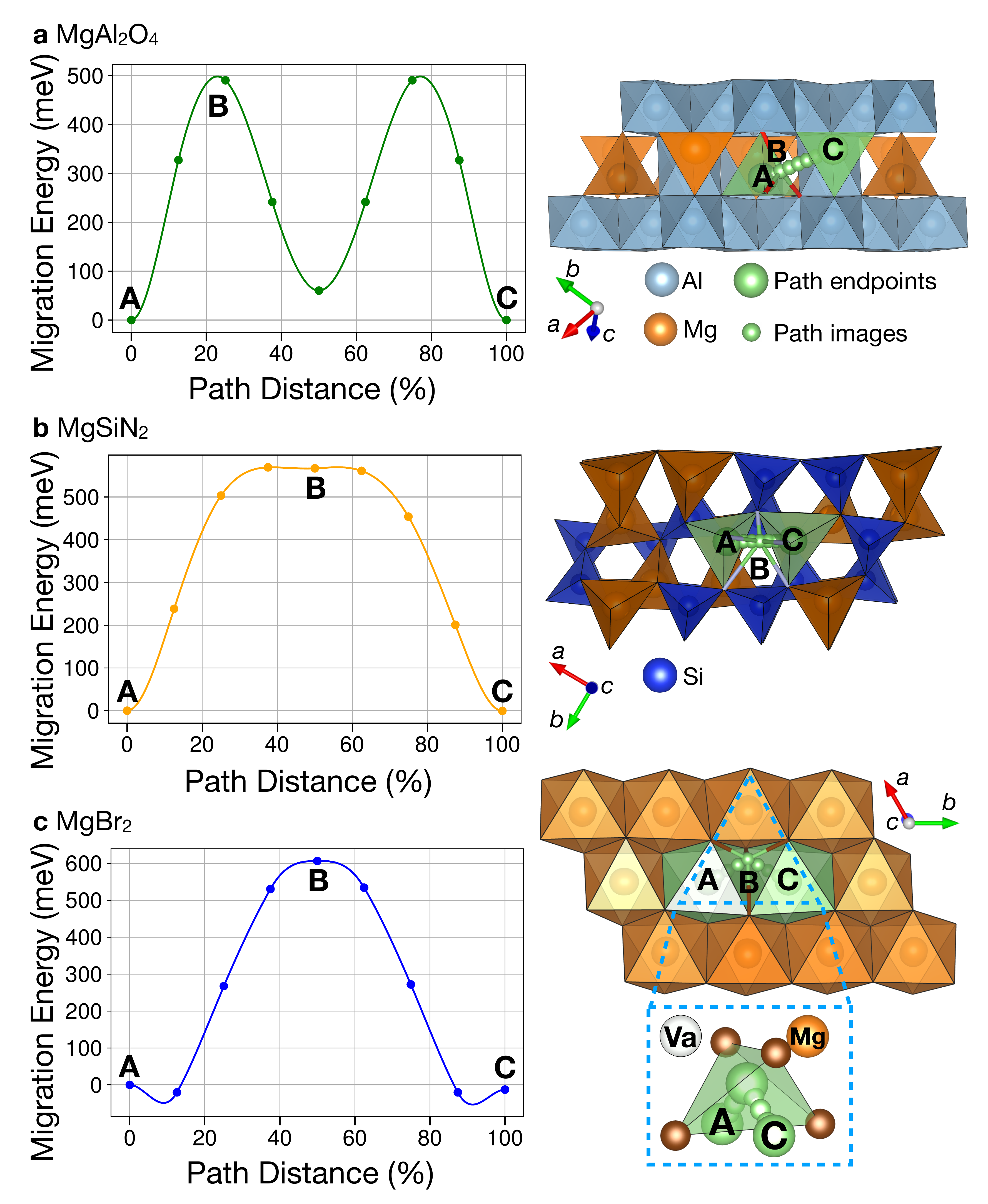}
\caption{Migration energies of {\bf a} MgAl$_2$O$_4$,  {\bf b} MgSiN$_2$, and {\bf c} MgBr$_2$ (left) along the calculated migration paths (right). For each migration path shown, the endpoints are labelled A and C (green polyhedra), the activated state is labelled B (green sphere with bonds), and the images along the path are shown as green spheres. Non-migrating Mg (orange), Al (light blue), and Si (dark blue), are shown as polyhedra with anions (not shown) at all polyhedral vertices. The inset in {\bf c} shows the nearest-neighbor octahedral cations (Va = vacuum) of the activated tetrahedral state.
\label{fig:mep}
}
\end{figure}

The evolution of the migration energies (left) along the calculated migration paths (right) of three promising materials, MgAl$_2$O$_4$  ({\bf a}), MgSiN$_2$  ({\bf b}), and MgBr$_2$  ({\bf c}), is displayed in Figure~\ref{fig:mep}. We chose to analyze MgAl$_2$O$_4$, MgSiN$_2$, and MgBr$_2$ owing to their low migration barriers ($<$~650~meV). Also, E$_m$ for MgAl$_2$O$_4$ and MgSiN$_2$ were calculated using step 4 in Figure~{\ref{fig:flow_chart}}{\bf a}, while MgBr$_2$ required steps 5-7. Non-migrating Mg, Al, and Si atoms in Figure~\ref{fig:mep} are indicated by orange, light blue, and dark blue spheres/polyhedra, respectively. The end-points, labelled ``A" and ``C" in both the migration energy plot and the migration paths, are shown as green polyhedra. The activated state that sets the migration barrier, labelled ``B", is shown with bonds to nearest-neighbor anions to identify its coordination environment. Anion atoms, which occupy all polyhedral vertices in Figure~\ref{fig:mep}, are not shown for clarity.

MgAl$_2$O$_4$ (Figure~\ref{fig:mep}{\bf a}) shows a migration energy landscape that is typical of spinel-oxides,\cite{Rong2015} where the stable tetrahedral sites (end-points) are connected via an intermediate octahedral site (not shown in Figure~\ref{fig:mep}{\bf a}). The activated site (B in Figure~{\ref{fig:mep}{\bf a}}) is the triangular face, i.e., migrating Mg coordinated to three nearby oxygen atoms, between the stable tetrahedral and the intermediate octahedral sites. Thus, Mg migration follows a ``4-3-6-3-4" pathway, where the numbers refer to the number of anions that are coordinated to the migrating Mg. E$_m$ for MgAl$_2$O$_4$ ($\sim$491~meV) is similar to the barriers observed, in the low vacancy limit, for oxide cathode spinels, such as MgMn$_2$O$_4$ (486~meV), MgCr$_2$O$_4$ (636~meV), MgCo$_2$O$_4$ (520~meV), and MgNi$_2$O$_4$ (485~meV),\cite{Liu2015} signifying possible compatibility (i.e., similar barriers + lattice) with cathode spinels. Interestingly, the intermediate octahedral site in MgAl$_2$O$_4$ is unstable by only $\sim$60~meV compared to the end-point tetrahedral sites and the low energy difference between the two sites may contribute to the low migration barrier.\cite{Canepa2017}

MgSiN$_2$ (Figure~\ref{fig:mep}{\bf b}) shows a migration path with stable tetrahedral sites and a 5-coordinated square pyramid as the activated site. Thus, Mg migrates via a 4-3-5-3-4 pathway, similar to topology of $\delta$-V$_2$O$_5$\cite{Gautam2015a,Gautam2015} and exhibits small coordination changes along the path, which may contribute to the low barrier (570~meV). Additionally, N$^{3-}$ in MgSiN$_2$ bonds with more covalently than O$^{2-}$ does, which can result in lower barriers via electrostatic shielding of the Mg$^{2+}$. Indeed, Mg-containing ternary Se$^{2-}$ and S$^{2-}$ spinels, which bond more covalently than O$^{2-}$, typically exhibit lower E$_m$ than analogous ternary oxide spinels.\cite{Canepa2017a} 

In Figure \ref{fig:mep}{\bf c}, MgBr$_2$ exhibits a barrier of $\sim$627~meV, where the Mg migrates across octahedral end-points through an activated, face-sharing tetrahedral site. Thus, the Mg migrates via a 6-3-4-3-6 mechanism, analogous to ionic migration in ordered, close-packed, layered oxides, such as Li in LiCoO$_2$\cite{Ven1999} and Mg in MgNiO$_2$.{\cite{Rong2015}} Note that both MgCl$_2$ and MgI$_2$ exhibit identical migration pathways compared to MgBr$_2$. Importantly, layered Mg-halides exhibit significantly lower migration barriers ($<$800~meV, Table~{\ref{tbl:summary}}) than MgNiO$_2$ ($>$1000~meV{\cite{Rong2015}}), which can be attributed to electrostatic interactions. For example, in MgNiO$_2$ (or analogous ordered, close-packed, layered oxides), the activated tetrahedron is face-shared with two non-migrating cation-occupied (Ni$^{3+/2+}$) octahedra, resulting in significant electrostatic destabilization, i.e. high energy due to strong electrostatic repulsion, and higher migration barriers. On the other hand, the activated tetrahedral site in layered Mg-halides (inset of Figure~{\ref{fig:mep}}{\bf c}), shares one face with a non-migrating Mg$^{2+}$-occupied octahedron (which contributes to electrostatic destabilization) and one face with the vacuum (Va) between the halide layers (which does not contribute to electrostatic destabilization), since layers of MgBr$_2$ are bonded via van der Waals interactions. Thus, the reduced electrostatic destabilization lowers the energy of the activated tetrahedral site and reduces the E$_m$ in MgBr$_2$ compared to MgNiO$_2$.  An alternative 6-2-6 mechanism, with Mg migrating through the shared edge of the MgBr$_6$ end-points can also be envisioned, but such a migration mechanism will require higher barriers, as previously demonstrated in MgNiO$_2${\cite{Rong2015}} and Mg$_2$Mo$_3$O$_8$.\cite{Gautam2016}

%%%%%%%%%%%%%%%%%%%%%%%%%%%%%%%%%%%%%%%%%%%%%%%%%%%%%%%%%%%%%%%%%%%%%%%%
\section{Discussion}
\label{sec:discussion}
%%%%%%%%%%%%%%%%%%%%%%%%%%%%%%%%%%%%%%%%%%%%%%%%%%%%%%%%%%%%%%%%%%%%%%%%

Using DFT calculations coupled with the NEB method, we have evaluated several binary, ternary, and quaternary compounds as potential anode/cathode coatings for high energy density Mg batteries. Protective coatings form an important component in the practical realization of Mg batteries due to the lack of stable (solid/liquid) electrolytes that are compatible with both high-voltage oxide cathodes and a Mg metal anode. Using a strict mobility threshold (600~meV, Figure~\ref{fig:thickness}) based on practical battery operating conditions, we identified MgSiN$_2$, Mg$_2$Ge, Mg$_3$As$_2$, and Mg$_3$P$_2$ as potential anode coatings and MgAl$_2$O$_4$ as a promising cathode coating (Figure~\ref{fig:barriers_bar}). MgB$_4$ also showed a low migration barrier (E$_m \sim$504~meV) but may not be suitable as an anode coating since it exhibits a small ESW ($\sim$~0.05~V) and is not stable against Mg metal (reductive stability of $\sim$~0.05~V). Additionally, we identified candidates that can function under more lenient operating conditions (980~meV), including MgH$_2$, Mg$_3$N$_2$, MgTe, MgSe, MgS, MgI$_2$, MgBr$_2$, and MgCl$_2$ on the anode and Mg(PO$_3$)$_2$ on the cathode. Finally, combining mobility and stability (ESW $>$1~V) requirements, we suggest MgSiN$_2$, MgSe, MgS, MgI$_2$, and MgBr$_2$ as promising anode coatings and MgAl$_2$O$_4$ and Mg(PO$_3$)$_2$ as possible cathode coatings.

The coatings with the highest oxidative stabilities in Figure 3 are MgF$_2$ (5.8~V), MgP$_4$O$_{11}$ (4.5~V), and MgS$_2$O$_7$ (4.5~V), which are also large gap ($>$6.5~eV) electronic insulators (Table~\ref{tbl:summary}). Particularly, MgF$_2$ is stable against Mg-metal and shows the largest ESW (5.8~V) among Mg-containing compounds.\cite{Chen2019} However, these materials cannot be considered as coatings due to their high Mg migration barriers (MgF$_2 \sim$1123~meV, MgP$_4$O$_{11} \sim$3541, and MgS$_2$O$_7 \sim$1699), which exceed even the lenient E$_{m}^{\mathrm{max}}$ threshold. Thus, our work highlights the importance of considering potential kinetic limitations when identifying coating candidates. Also, we have restricted our work here to structures involving a unique anion, (i.e., halides, chalcogenides, pnictides, etc.), but it would be interesting to evaluate Mg-compounds/structures with mixed anions (e.g., oxy-sulfides, oxy-halides, oxy-nitrides, etc.) for coating applications since having multiple anions will provide an additional handle to calibrate the stability, mobility, and electronic properties. 

One important assumption made in this work is that long-range ionic diffusion in coating materials follows a random-walk model without any appreciable degree of correlation between local migration events, i.e., $f \sim 1$ in Eq.~\ref{eq:diffusivity}. While several cathodes and solid electrolytes indeed obey random-walk diffusion, as indicated by the robust agreement between experimentally measured and theoretically predicted migration barriers, \cite{Meutzner2018} it remains to be seen if any degree of correlation exists in coating materials. Computationally expensive AIMD or kinetic Monte-Carlo simulations can provide some theoretical evidence for correlation during diffusion (if it exists), while experimental measurements, such as impedance spectroscopy (IS) and/or nuclear magnetic resonance (NMR), can also be used to probe correlation. Further, we assume that diffusion carriers are already present in the coating materials during battery operation in Eq.~1, either via transfer of Mg at the coating$||$electrode interface or due to intrinsic defects. If the formation of diffusion carriers has a significant energy cost (E$_f$), then the Mg diffusivity will drop by $\exp{(-\frac{{\rm E}_f}{k_BT})}$ compared to estimates using only E$_m$ (Eq.~\ref{eq:diffusivity}).

We found that the layered MgX$_2$ structures (X~=~Cl, Br, and I) exhibit significant structural distortion upon addition of a jellium background, which is typically used for charge-compensation in periodic boundary DFT calculations. To circumvent this limitation, we devised a three-step procedure (steps 5-7, Figure~{\ref{fig:flow_chart}}{\bf a}) to calculate Mg migration barriers, which can result in an error of $\sim$160~meV (Figure~{\ref{fig:flow_chart}}{\bf c}). While the DFT+NEB workflow (Figure~{\ref{fig:flow_chart}}{\bf a}) can be generalized to evaluate ionic transport in other material chemistries as well, more theoretical studies are required to further develop alternate strategies to model local charge imbalances and to better quantify errors. In particular, experimental measurements of Mg migration barriers, using variable temperature (VT) IS or VT-NMR, would be useful to validate and improve the theoretical description of candidate coating materials.

Analyzing the Mg migration topology of three distinct candidates with E$_m <$650~meV (Figure~{\ref{fig:mep}}), we found similarities between MgAl$_2$O$_4$ and oxide cathode spinels, MgSiN$_2$ and $\delta-$V$_2$O$_5$, and MgBr$_2$ and close-packed layered oxide cathodes. In particular, MgAl$_2$O$_4$ should exhibit low lattice mismatch with oxide spinel cathodes, such as spinel-Mg$_{\rm x}$Mn$_2$O$_4$, and also has a similar E$_m$ at the low vacancy limit, highlighting its suitability as a cathode coating. However, MgAl$_2$O$_4$ against Mg-metal is likely to decompose further into Al and MgO{\cite{Chen2019}} and is thus unlikely to be suitable as an anode coating.

On the other hand, MgSiN$_2$ (Figure~\ref{fig:mep}{\bf b}) is stable against Mg metal (hence a potential anode coating) but shows a low oxidative stability ($\sim$1.2~V) and may not be compatible with low voltage sulfide cathodes (e.g., Mg$_x$Mo$_6$S$_8$ and Mg$_x$Ti$_2$S$_4$ \cite{Aurbach2000, Sun2016}) due to differences in structure and anion (N$^{3-}$ vs.~S$^{2-}$). The low E$_m$ in MgSiN$_2$ can be attributed to the small changes in the coordination environment along the migration pathway as well as the covalent bonding exhibited by N$^{3-}$ compared to O$^{2-}$.  The recent demonstration that theoretically predicted novel ternary nitride compounds can be experimentally synthesized{\cite{Sun2019}} holds promise in the identification of new Mg-containing nitrides with appreciable thermodynamic stability and Mg mobility to function as coating materials. 

Due to the lack of atomic occupation in the interlayer spacing, layered Mg-halides, i.e. MgCl$_2$, MgBr$_2$ (Figure~\ref{fig:mep}{\bf c}), and MgI$_2$, exhibit an interesting structural motif not usually observed in other Mg-compounds. While close-packed layered compounds (e.g., MgNiO$_2$) are typically expected to yield larger Mg migration barriers due to strong electrostatic interactions and the octahedral coordination preference of Mg,\cite{Rong2015} the lack of electrostatic destabilization of the activated tetrahedral site significantly lowers the barrier in MgBr$_2$. Such reduction in migration barriers, by lowering electrostatic interactions, has been demonstrated in disordered, close-packed, layered, Li-excess oxide cathodes.\cite{Lee2014} Thus, layered/close-packed structures that can lower the electrostatic destabilization of the activated site, via cation disorder for example, might be a promising motif to identify novel materials with facile Mg mobility. 

Although there is no broad association between Mg migration barriers and electronic band gaps, barriers do decrease monotonically with decreasing band gaps within an anion group, likely due to trends in volume per anion and atomic orbital overlap (Table~{\ref{tbl:summary}}). In general, higher electronic conductivity in coatings compared to electrolytes is detrimental to electrolyte stability and can cause battery self-discharge.{\cite{Nakamura2019}} Two candidates with relatively low Mg$^{2+}$ migration barriers ($<$600~meV), namely MgB$_4$ (band gap $\sim$0.03 eV) and Mg$_2$Ge ($\sim$0.61~eV), have a small band gap ($<$1~eV, high electronic conductivity) and small EWS ($<$1~V, low ability to accommodate a large chemical potential difference). Hence, MgB$_4$ and Mg$_2$Ge will conduct both electrons and Mg$^{2+}$, causing electrolyte instability, and are not suitable as coating materials. There are also candidates with sufficient Mg mobility and ESW ($>$1~V) but with low band gaps ($<$2~eV), such as MgSiN$_2$, MgSe, and MgI$_2$, which can result in higher electronic conductivities within the coating than the electrolyte. However, using thicker layers of MgSiN$_2$, MgSe, and MgI$_2$ as anode coatings may mitigate the influence of electronic conductivity and accommodate electrolytes with narrow ESWs.

%The theoretical framework used in this study can be readily extended to other mono or multivalent chemistries to search for novel coating materials. In the case of Mg batteries, we have restricted our work here to structures involving a unique anion, i.e., halides, chalcogenides, pnictides, etc.), but it may be interesting to evaluate the thermodynamic stability and Mg mobility of compounds/structures with mixed anions (e.g., oxy-sulfides, oxy-halides, oxy-nitrides, etc.) as multiple anions will provide a complete blueprint of the the stability, mobility, and electronic properties in these materials. 

Ideally, coatings in Mg batteries should: \emph{i}) be inexpensive, \emph{ii}) involve simple equipments, and \emph{iii}) not alter the composition/properties of electrodes and electrolytes. Several strategies exist to introduce coating materials onto electrodes in Li-ion batteries, as summarized recently by Culver \emph{et al.}{\cite{Culver2019}} Inexpensive and simpler techniques include wet and chemical spray coating, while expensive and advanced techniques include ALD, PLD, and chemical vapor deposition. In general, the cost of a coating increases dramatically if thinner (sub-nanometer) layers are to be formed.{\cite{Culver2019}} It remains to be seen if coating techniques routinely applied to Li-ion batteries can be translated directly to Mg (and multivalent) batteries. In particular, the low tolerance of Mg battery components toward oxygen contamination may represent a significant challenge in the preparation of \emph{ex situ} coated electrodes.

%%%%%%%%%%%%%%%%%%%%%%%%%%%%%%%%%%%%%%%%%%%%%%%%%%%%%%%%%%%%%%%%%%%%%%%%
\section{Conclusion}
\label{sec:conclusion}
%%%%%%%%%%%%%%%%%%%%%%%%%%%%%%%%%%%%%%%%%%%%%%%%%%%%%%%%%%%%%%%%%%%%%%%%

In this study, we use density functional theory calculations to identify protective coating materials for Mg batteries, a potential, beyond Li-ion, high energy density secondary electrochemical storage system. Based on a set of minimal ionic mobility requirements in potential coatings, which are applicable to all intercalation battery systems, we found a wide variety of candidate coating materials. For example, using a strict mobility threshold (600~meV), we identified MgSiN$_2$, MgB$_4$, Mg$_2$Ge, Mg$_3$As$_2$, Mg$_3$P$_2$, and MgAl$_2$O$_4$ as potential coatings, while using a lenient threshold (980~meV) extended our candidate set to include MgH$_2$, Mg$_3$N$_2$, MgTe, MgSe, MgS, MgI$_2$, MgBr$_2$, MgCl$_2$, and Mg(PO$_3$)$_2$. Amongst the aforementioned candidates, MgAl$_2$O$_4$ and Mg(PO$_3$)$_2$ should be compatible against oxide cathodes ($>$3~V vs.~Mg), while MgSiN$_2$, MgS, MgSe, MgBr$_2$, and MgI$_2$ should be compatible with the Mg metal anode and exhibit reasonable ESW ($>$1~V). Upon inspecting the Mg migration topology in a subset of the candidates listed above, we observed similarities with other migration pathways, such as spinels, $\delta-$V$_2$O$_5$, and close-packed layered structures. Further, we expect that layered frameworks, similar to MgX$_2$ (X~=~Cl, Br, and I), can exhibit reasonable Mg migration barriers due to the lower electrostatic destabilization from the vacuum interlayer that face-shares with the activated site during migration. Finally, we suggest careful and dedicated experimental measurements combined with advanced characterization techniques to further validate our theoretical predictions and progress towards practical Mg batteries.

%%%%%%%%%%%%%%%%%%%%%%%%%%%%%%%%%%%%%%%%%%%%%%%%%%%%%%%%%%%%%%%%%%%%%%%% 
%% The "Acknowledgement" section can be given in all manuscript
%% classes.  This should be given within the "acknowledgement"
%% environment, which will make the correct section or running title.
%%%%%%%%%%%%%%%%%%%%%%%%%%%%%%%%%%%%%%%%%%%%%%%%%%%%%%%%%%%%%%%%%%%%%%%%

\section*{Acknowledgements}
P.\ C. acknowledges support from the Singapore Ministry of Education Academic Fund Tier 1 (R-284-000-186-133).  T.\ C. is supported by the National Science Foundation Graduate Research Fellowship under Grant No.\ DGE 1106400. The computational work for this article was performed on resources of the National Supercomputing Centre, Singapore (https://www.nscc.sg) and the National Science Foundations Extreme Science and Engineering Development Environment (XSEDE) supercomputer Stampede2, through allocation TG-DMR970008S, which is supported by National Science Foundation grant number ACI-1548562.

%%%%%%%%%%%%%%%%%%%%%%%%%%%%%%%%%%%%%%%%%%%%%%%%%%%%%%%%%%%%%%%%%%%%%%%%

%%%%%%%%%%%%%%%%%%%%%%%%%%%%%%%%%%%%%%%%%%%%%%%%%%%%%%%%%%%%%%%%%%%%%%%%
%% The same is true for Supporting Information, which should use the
%% suppinfo environment.
%%%%%%%%%%%%%%%%%%%%%%%%%%%%%%%%%%%%%%%%%%%%%%%%%%%%%%%%%%%%%%%%%%%%%%%%
%%%%%%%%%%%%%%%%%%%%%%%%%%%%%%%%%%%%%%%%%%%%%%%%%%%%%%%%%%%%%%%%%%%%%%%%
\begin{suppinfo}
The Supporting Information is available free of charge on the ACS Publications website at DOI: 
All the migration barriers and their migration paths for all compounds are reported.
\end{suppinfo}

%%%%%%%%%%%%%%%%%%%%%%%%%%%%%%%%%%%%%%%%%%%%%%%%%%%%%%%%%%%%%%%%%%%%%%%%
%% TOC
%%%%%%%%%%%%%%%%%%%%%%%%%%%%%%%%%%%%%%%%%%%%%%%%%%%%%%%%%%%%%%%%%%%%%%%%
%\begin{tocentry} 
%\includegraphics{toc} 
%\end{tocentry}

%%%%%%%%%%%%%%%%%%%%%%%%%%%%%%%%%%%%%%%%%%%%%%%%%%%%%%%%%%%%%%%%%%%%%%%%
% References and Notes
\bibliography{mg_transport_coatings}
%%%%%%%%%%%%%%%%%%%%%%%%%%%%%%%%%%%%%%%%%%%%%%%%%%%%%%%%%%%%%%%%%%%%%%%%

%%%%%%%%%%%%%%%%%%%%%%%%%%%%%%%%%%%%%%%%%%%%%%%%%%%%%%%%%%%%%%%%%%%%%%%%
\end{document}